\documentclass[11pt,letterpaper]{article}
\usepackage[utf8]{inputenc}
\usepackage{url}
\usepackage{amsmath} 
\usepackage{graphicx}
\usepackage{amsfonts}
\usepackage{epstopdf}
\usepackage{amsopn}
\date{}

\usepackage[parfill]{parskip}

\title{Dynamical-System Model Predicts\\ When Social Learners Impair Collective Performance}

\author{Vicky Chuqiao Yang\thanks{Santa Fe institute, Santa Fe, NM 87501}
\and Mirta Galesic\thanks{Santa Fe institute, Santa Fe, NM 87501; Complexity Science Hub Vienna, Austria; Vermont Complex Systems Center, University of Vermont, VT 05405 }
\and Harvey McGuinness\thanks{Zanvyl Krieger School of Arts and Sciences, Johns Hopkins University, Baltimore, MD 21218}
\and Ani Harutyunyan\thanks{Sunwater Institute, North Bethesda, MD 20852}}

\begin{document}

\maketitle

\begin{abstract}
A key question concerning collective decisions is whether a social system can settle on the best available option when some members learn from others instead of evaluating the options on their own. This question is challenging to study, and previous research has reached mixed conclusions, because collective decision outcomes depend on the insufficiently understood complex system of cognitive strategies, task properties, and social influence processes. This study integrates these complex interactions together in one general yet partially analytically tractable mathematical framework using a dynamical system model. In particular, it investigates how the interplay of the proportion of social learners, the relative merit of options, and the type of conformity response affect collective decision outcomes in a binary choice. The model predicts that when the proportion of social learners exceeds a critical threshold, a bi-stable state appears in which the majority can end up favoring either the higher- or lower-merit option, depending on fluctuations and initial conditions. Below this threshold, the high-merit option is chosen by the majority. The critical threshold is determined by the conformity response function and the relative merits of the two options. The study helps reconcile disagreements about the effect of social learners on collective performance and proposes a mathematical framework that can be readily adapted to extensions investigating a wider variety of dynamics.
\end{abstract}
\section{Introduction}
Collective decisions are central to human societies, from small-scale social systems such as families and committees to large-scale ones such as democratic governments and international organizations. Some of the most pressing challenges facing humanity, for instance, addressing climate change, global pandemics, and economic inequalities,  critically depend on collective decisions. A central concern regarding these systems is the effect of social learners, here defined as individuals who adopt other people’s opinions and behaviors rather than explore the merit of available options on their own. This study investigates how social learners affect collective performance.

Previous research on the effect of social learning on collective decision outcomes has come to mixed conclusions. Some find social learners impair collective performance \cite{lorenz2011social, rogers1988does}, some find them beneficial \cite{jayles2017social, couzin2011uninformed}, and some argue they depend on the network structure, adaptability, or the level of network effects \cite{almaatouq2020adaptive, becker2017network, gray2018multiagent, brock2001discrete}. The question is challenging to address because collective decision outcomes depend on the insufficiently understood interactions of multiple cognitive and social factors, including cognitive strategies relying on individual or social learning, task properties, and the social influence processes. Few attempts have been made at developing overarching mathematical frameworks capable of integrating these complex interactions into parsimonious theories. 

Most previous research focuses on tasks where one option clearly has more merit than others (e.g., \cite{becker2017network,lorenz2011social,jayles2017social,jayles2020impact,simoiu2019studying}). These studies have found that social learning can impair collective performance \cite{rogers1988does}, although adaptive learners who can switch from social to individual learning can help avoid this fate \cite{enquist2007critical,rendell2010rogers}, especially if more-accurate individuals are less likely to copy others \cite{becker2017network} and if more-accurate individuals are preferentially copied \cite{almaatouq2020adaptive}. Other research is less concerned about the relative merit of options and instead focuses on investigating factors influencing group consensus such as the initial proportion of individuals supporting each option, their position in the group, and the strength of their preference for that option \cite{conradt2009leading,dyer2009leadership,couzin2011uninformed}. Few theoretical frameworks are able to contain both scenarios.

Furthermore, most studies of collective intelligence do not explicitly compare different ways in which social learners can be influenced by others. Different shapes of conformity functions have been observed in human groups. In some cases, social learners adopt a belief or behavior with a likelihood that is higher than its observed frequency, exhibiting an s-shape relationship with the observed frequency (similar to the solid blue curve in Fig.~\ref{fig:modelCartoon}(B)). It has been hypothesized that this kind of conformity function is associated with normative conformity, whereby individuals are concerned primarily with fitting into a group \cite{claidiere2012integrating}. In other cases, the likelihood of adopting a behavior or belief still increases with its observed frequency, but at a linear or slower pace, resulting in a linear or an inverse s-shape relationship with the observed frequency (similar to the dashed orange curve in Fig.~\ref{fig:modelCartoon}(B)). This function has been hypothesized to be associated with informational conformity, where the primary motivation for social learning is finding the best available option \cite{claidiere2012integrating}. While the functional form that best describes different types of conformity is an empirical question, here we use the two conformity curves to approximate the distinction between these two types of conformity, which are rarely systematically compared in studies of collective decisions.

In this paper, we integrate and explicitly compare these different assumptions within an overarching mathematical framework that enables exploration of the dynamic complex system underlying collective performance, composed of cognitive strategies (individual vs.~social learning), task properties (relative merit of options), and social influence processes (normative vs.~informational conformity). We show how the framework helps to understand sometimes contradictory findings of the effect of social learners on the quality of collective decisions. 

\section{Summary of the Mathematical Model}
\begin{figure*}[htb] 
    \centering
    \includegraphics[width = 0.8\linewidth]{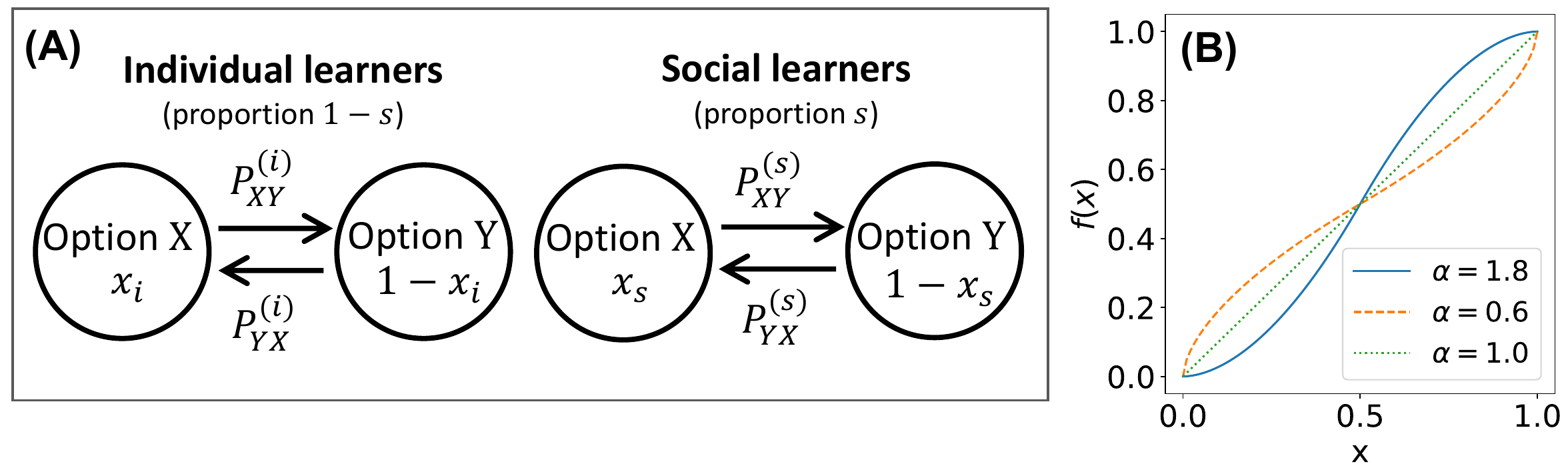}
    \caption{(A) Sketch of the model setup showing the interpretation of parameters. (B) The conformity function for a few shape parameter $\alpha$ values, leading to different concavities of the function ($\alpha>1$ has been associated with normative conformity and $\alpha<1$ with informational conformity). The vertical axis, $f(x)$, shows the transition rates from option $Y$ to option $X$. The horizontal axis, $x$, shows the observed frequency of option $X$ in the population.}
    \label{fig:modelCartoon}
\end{figure*}

We consider options $X$ and $Y$ to compete for followers in a well-mixed population consisted of social and individual learners, where the proportion of social learners is $s$. We denote the merit of $X$ relative to $Y$ to be $m$, where $m$ is a number between 0 and 1. For each type of cognitive strategy, we use the conservation relationship described in the group competition framework \cite{abrams2003linguistics}---the change in the proportion of population favoring an option equals those switching to the option minus those switching to the other option (see Fig.~\ref{fig:modelCartoon}(A) for an illustration). Mathematically, this is expressed as
\begin{equation}\label{eqm:1}
    \frac{dx_j}{dt} = P_{YX}^{(j)} (1-x_j) - P_{XY}^{(j)}  x_j \;,\\
\end{equation}

where $j$ denotes the cognitive strategy under consideration, $j = i$ denotes variables for individual learners, and $j = s$ denotes those for social learners. The term $x_j$ is the proportion of people favoring option $X$ among those using cognitive strategy $j$, $P_{YX}^{(j)}$ denotes the transition rate from $Y$ to $X$, and $P_{XY}^{(j)} $ is the transition rate vice-versa. In this model, we study the interplay of three main elements of the complex social system underlying collective decisions: cognitive strategy, the merit of different options, and conformity response. We formulate the transition rates of social and individual learners based on different mechanisms. For individual learners, the transition rates depend only on the merit of $X$ relative to $Y$, denoted as $m$. The merit is the proportion of individuals who would choose an option after fully evaluating all available information individually, such as assessing factual evidence or one's own moral values. $m$ is a number between $0$ and $1$, and $m = 0.5$ denotes the two options having equal merit. For social learners, the transition rates depend on the proportion of people favoring option $X$, and the conformity response function, $f(x)$, whose shape is parameterized by one parameter, $\alpha$. Altering $\alpha$ leads to functional forms associated with either informational or normative conformity (see Fig.~\ref{fig:modelCartoon}(B) for a few examples). After deriving the transition rates for the two types of learners, we analyze the fixed points of Eq.~\ref{eqm:1} and their stability. Please see SI 1--3 for more details on the model's derivation and solutions.

\section{Results}

\begin{figure*}[htb]
	\centering
	\includegraphics[width = \linewidth]{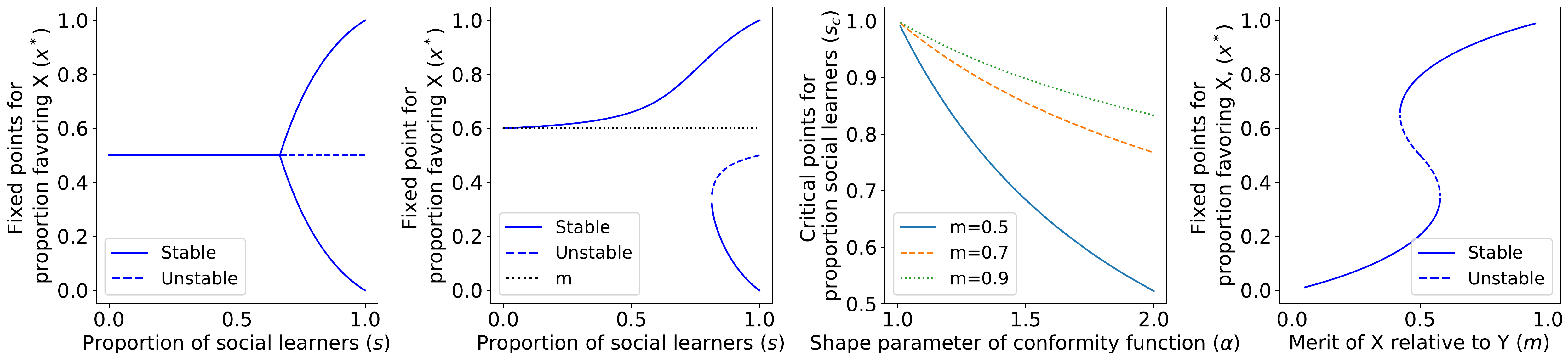}
	\caption{The model's predictions in relation to parameters $s$, $\alpha$, and $m$. (A) The fixed points of the proportion of people favoring option $X$, $x^*$, as a function of the proportion of social learners, $s$, when the two options have equal merit, $m = 0.5$. For $s$ greater than a critical value, a bi-stable state appears. (B) The same analysis as in (A) but for the case of unequal merit, $m = 0.6$. (C) Critical transition point for $s$, $s_c$, as a function of $\alpha$ for a few $m$ values. The area above the curves represents parameter regimes where bi-stable states appear. (D) The fixed points' dependency on $m$ when $s = 0.8$. They exhibit a bi-stable state for mid-range $m$, and the system can display path dependency if $m$ changes. In (A), (B) and (D), $\alpha = 1.5$. 
	}
	\label{fig:result_combo}
\end{figure*}

We study the model's predictions by analyzing the fixed points of Eq.~\ref{eqm:1} for $x$ (denoted as $x^*$) and their stability. The fixed points signify where $x$ stops changing over time. A fixed point is stable if the system, when perturbed, returns to that point, and is unstable if the system departs from this fixed point when perturbed.

The model predicts that the proportion of social learners, $s$, significantly affects the system's behavior. When social learners adopt an option with a likelihood higher than its frequency ($\alpha >1$, associated with normative conformity), our model predicts that the majority of the population will choose the option with higher merit, but only if the proportion of social learners is below a critical threshold. When the proportion of social learners exceeds the critical threshold, a bi-stable state appears in which the majority can favor either the higher- or lower-merit option, depending on fluctuations and initial conditions. Fig.~\ref{fig:result_combo}(A) shows the dependency of $x^*$ on $s$ for $m = 0.5$ ($X$ and $Y$ have equal merit). The bi-stable state occurs when $s$ is above a threshold $s_c$. At $s_c$, the dynamical system undergoes a pitchfork bifurcation, where the stability of the fixed point $x=0.5$ changes. In Fig.~\ref{fig:result_combo}(B), we show results for a case of unequal merits ($m = 0.6$, $X$ has higher merit than $Y$), and a bi-stable state is also present, while the symmetry of the bifurcation is broken. 

The critical threshold $s_c$ is determined by parameters $\alpha$ and $m$. In Fig.~\ref{fig:result_combo}(C), we present numerical solutions for $s_c$ for a number of $m$ values as a function of $\alpha$. The region of $s>s_c$ is where the bi-stable state occurs. The critical point $s_c$ decreases with $\alpha$, and also decreases as $m$ gets closer to 0.5. Note that in the bi-stable region, if $m$ also evolves, the system is path dependent. As shown in Fig.~\ref{fig:result_combo}(D), if the merit $m$ increases from a small value, the stable fixed point $x^*$ first increases along the lower branch, and after $m$ grows past the critical point, $x^*$ goes through a sudden jump to the upper branch. If $m$ is again lowered, $x^*$ does not go through the same jump to the lower branch. Instead, $x^*$ stays on the upper branch until $m$ drops below the other bifurcation point. The existence of path dependence suggests that when social learning is strong, beliefs and behaviors persist over time despite changes in the merit of options. It is also possible for the the proportion of individuals who favor one option to jump suddenly after sufficient change in merit. 

The occurrence of a bi-stable region is not a result of the assumed specific functional form of $f(x)$. We generalize our model by considering general increasing $f(x)$ symmetrical about the midpoint, $(0.5, 0.5)$. In the case of equal merit, we show analytically that the bi-stable region occurs for $s \geq s_c = {1}/{f'(0.5)} $. For derivation, see SI 3. This result also predicts that bi-stable state exists only when $f'(0.5)>1$, that is, when the social learners have a superlinear response to the frequency of options near a 50-50 split. This result also explains why the bi-stable state does not appear in the $\alpha \leq 1$ case (see SI 4 for model behavior under this condition). 

Besides the well-mixed model with two types of learners (the baseline model), we consider additional extensions. The first extension relaxes the assumption of a dichotomy between individual and social learners, allowing individuals to adopt learning strategies that are on a spectrum of social and individual learning (SI 5.1). The second extension considers individuals connected on a 2-D lattice, where social learners are affected only by their local neighborhood (SI 5.2). The third extension considers the introduction of noise in the system (SI 5.3). In all extensions, we find that our major conclusions regarding the critical threshold of social learners remain unchanged. 

\begin{figure}[htb] 
	\centering
	\includegraphics[width = 0.7\linewidth]{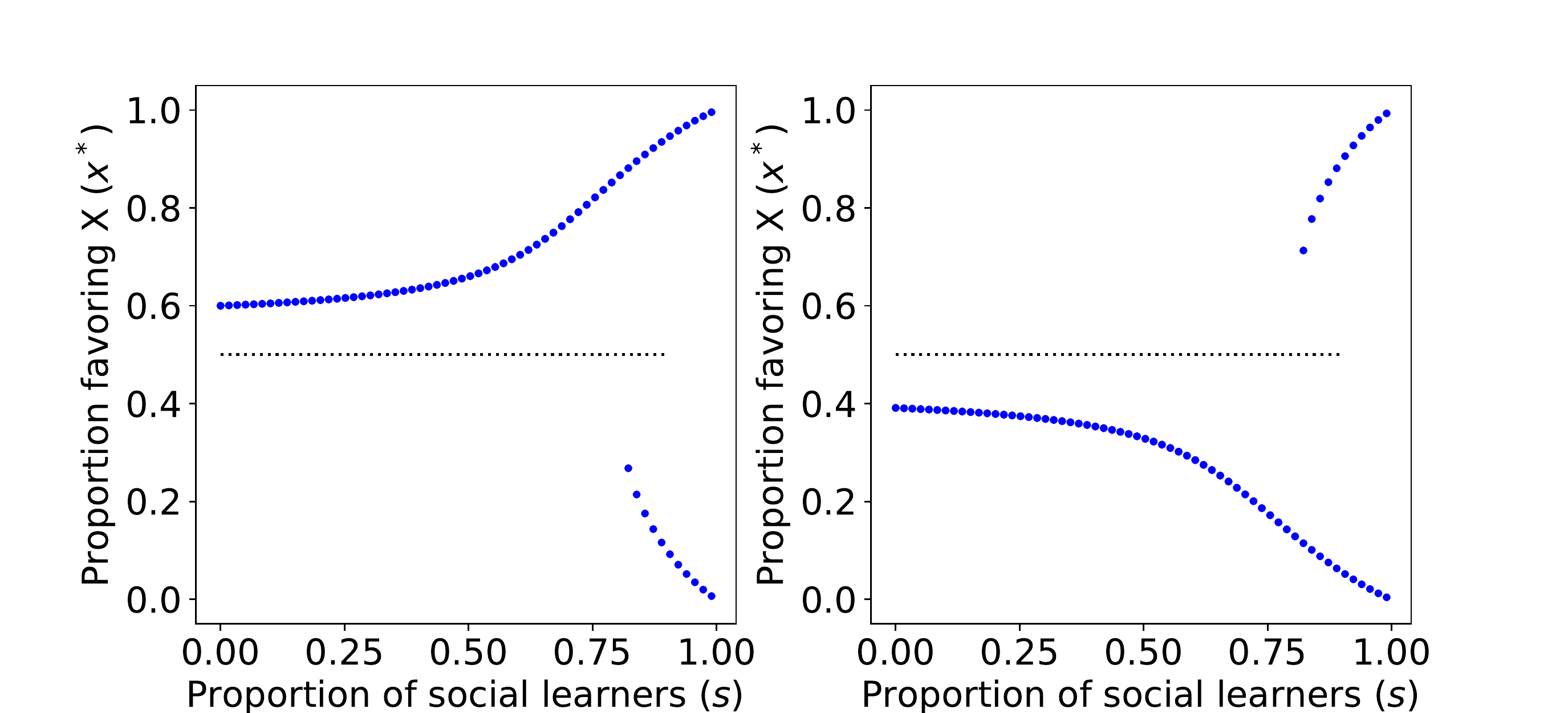}
	\caption{Simulation results of the opinion strength extension model. The proportion of individuals favoring option $X$ as a function of the proportion of social learners. In both panels, $m = 0.6$, assigning $X$ higher merit than $Y$, so that the desired outcome is $x^*> 0.5$. The black dotted line shows $x^* = 0.5$ for reference. (A) Two options are held with equal strength. The results are similar to that of the baseline model (Fig.~\ref{fig:result_combo}(B)). (B) The low-merit option ($Y$) is held with greater strength. The two branches of the bifurcation are flipped, and the model recovers the finding in Couzin et al. \cite{couzin2011uninformed}. $X$ is analogous to the option weakly preferred by the majority; $Y$ is analogous to that strongly preferred by the minority. In both panels, the parameter $\alpha = 1.5$.}
	\label{fig:strength_extension}
\end{figure}

Next, we consider another extension where we introduce an additional variable for the strength of opinions about options (opinion strength extension), leading to more complex dynamics. This extension is motivated by a model developed by Couzin et al.~\cite{couzin2011uninformed}, which considers an animal group consisting of informed and uninformed individuals and reaches seemingly contradictory predictions. The uninformed and informed individuals act similarly to the social and individual learners in our model, respectively, with some differences (see SI 5.4 for details). Couzin et al.~particularly consider the scenario where individuals holding the minority preference (similar to the low-merit option in our model) are more stubborn and less likely to update their views. They find that when the uninformed individuals are below a certain threshold, the minority preference is more likely to prevail. When the proportion of uninformed individuals is above that threshold, a bi-stable state occurs, and the likelihood for the majority preference to prevail increases. 

We add the strength of opinion as an additional variable by extending the transition rates of the individual learners from $X$ to $Y$ to be  $P_{XY}^{(i)} = (1-w_x) (1-m)$, where $w_x$, a parameter taking values between $0$ and $1$, is the relative strength of opinion about option $X$ compared to that about $Y$, and $w_x = 0.5$ denotes equal strength. This modification considers a probability $w_x$ that an individual learner holding option $X$ stays with this option, regardless of merit. Similarly, the transition rates of individual learners from $Y$ to $X$ is $P_{YX}^{(i)} = w_x m$. The transition rates of the social learners and the other dynamics of the system remain the same. When we simulate this extension with the two options held with equal strength, the results are shown in Fig.~\ref{fig:strength_extension}(A), and we recover the predictions of the baseline model (Fig.~\ref{fig:result_combo}(B)) as expected. Because the analysis is through simulations, only stable fixed points are shown. When we simulate the case where the low-merit option is held with greater strength, the two branches of the bifurcation are flipped, and the model recovers the finding in Couzin et al. Please see SI 5.4 for more details of this extension and the comparison between the two models.  

\section{Discussion}

\begin{table*}[htb]
\centering
\begin{tabular}{ p{7cm}|p{4cm}|p{4.4cm}  }
  \hline
&  $s < s_c$ \newline (fewer social learners)  &  $s > s_c$ \newline (more social learners) \\ 
  \hline
  \hline
$\alpha > 1$ \newline (s-shaped function, associated with normative conformity) & high-merit & high- or low-merit, depends on initial conditions and fluctuations\\
 \hline
$\alpha \leq 1$ \newline (linear or inverse s-shaped function, associated with informational conformity) & \multicolumn{2}{|c}{high-merit } \\
   \hline
\end{tabular}
\caption{Summary of the model's prediction of collectively chosen options under majority rule.}
\label{tab:res_summary}
\end{table*}
We study the joint effects of the proportion of social learners, the relative merit of different options, and processes of social influence on the collective decision outcome. Our baseline model suggests that when social learners copy an option with a likelihood higher than its frequency (a process associated with normative conformity), their effect on the collective decisions depends critically on their proportion in the system. The model predicts a threshold for the proportion of social learners, below which the high-merit option would be preferred by the majority. When the proportion of social learners exceeds this threshold, it becomes possible for the low-merit option to be favored by the majority. The model also predicts that when the proportion of social learners is above the critical threshold, the system is path dependent if $m$ changes over time---it can be hard to reverse the change in the proportion of individuals favoring $X$ if $m$ reverses. However, if the social learners copy an option with the likelihood that is equal or smaller than its frequency (a process associated with informational conformity), the critical threshold ceases to exist, and the majority always favors the high-merit option. 

Our model is formulated in terms of the proportion of individuals in a system, which would work best when the group size is large. The same mechanism can also be adapted to study small groups. For example, both the spectrum extension and the spatial extension of our model, presented in SI 5.1 and 5.2, respectively, are formulated on the individual level and can apply to groups of finite size. 

Our model offers several contributions to the literature. First, the model characterizes collective decision-making problems using three parameters, the combination of which distinguishes a broad range of group-level behaviors (see Table~\ref{tab:res_summary} for a summary). These parameters describe cognitive strategies ($s$), the problem at hand ($m$), and the conformity response ($\alpha$). The model predicts a threshold (that depends on $m$ and $\alpha$) above which social learners hurt the collective decision outcomes, and under which they do not, suggesting that the same model can explain seemingly conflicting findings about collective performance. Second, the model is analytically tractable, such as in drawing conclusions for the behavior of the system for general $f(x)$ detailed in SI 3. This expands the simulation studies typical in studying collective decisions and enables us to draw conclusions for a broad class of behaviors. Third, our model offers a flexible mathematical framework that can be extended or adapted to study interactions of a wide range of social and psychological factors for future research, as demonstrated in the extensions outlined in SI 5. 

The problem of how social learners influence collective performance is highly relevant to democratic elections. It has been noted that a sizable proportion of the voting population is uninformed about options they are voting on, such as policies proposed by the candidates \cite{somin2016}. These uninformed voters might instead use social learning to make decisions \cite{buchanan1962calculus}. While there is a long-standing consensus on the widespread phenomenon of such uninformed voters \cite{converse2006nature,bartels1996uninformed}, there is little agreement on how they affect the outcome of democratic elections \cite{somin2016}. On the one hand, some argue that the presence of uninformed voters does not affect election outcomes, because information shortcuts such as political parties and opinion leaders are sufficient to lead to good decisions, or the uninformed decisions can be taken as noise that cancels out \cite{bartels1996uninformed, bhatti2010would}. On the other hand, some argue that this optimistic view is at odds with the empirical evidence \cite{grosser2016curse, bartels1996uninformed} and that uninformed voters can lead to adoption of bad policies \cite{caplan2011myth}. Our baseline model suggests that both outcomes can occur depending on conditions, in particular, whether the proportion of uninformed voters is above or below the critical threshold determined by the conformity response and the relative merit of the two options. Further, our baseline model predicts the possibility of sudden shifts in collective decision outcomes when the proportion of uninformed voters is above the critical threshold. In addition to the predictions of the baseline model, the results from the opinion strength extension suggest that the role of social learners may differ depending on the presence of committed opinion holders. It is in line with findings from previous research---both theoretical and experimental studies have found that the presence of inflexible opinion holders significantly alters the outcomes of the collective \cite{xie2011social, centola2018experimental, mobilia2007role, galam2007role}. Ultimately, if an individual learner never changes their mind based on new evidence, their chosen option could spread throughout the social network \cite{mobilia2003does,mobilia2007role}. This suggests committed minorities can have an important influence over the outcomes of democratic elections.

This paper offers insights into the complex process of collective decision making through a parsimonious model. We show that our main conclusions are robust to several relaxations of the model assumptions, such as considering that individuals can fall on a spectrum between pure social and pure individual learning or that the social influence is only local (SI 5). We aim to provide a simple framework that can integrate parallel efforts in different disciplines studying collective behavior. Starting from cognitive and social principles, our group-level framework recovers bi-stable states predicted by individual-level models \cite{bizyaeva2020general, couzin2011uninformed, brock2001discrete}. Besides its benefit of simplicity, the group-level framework could help recognize when apparently different mechanisms (e.g., stubbornness and relative merit of different options in our opinion strength extension) affect group-level outcomes in the same direction. Our work also suggests several directions for future research. While the S-shaped and inverse S-shaped curves were hypothesized to be associated with normative and informational conformity \cite{claidiere2012integrating}, more research is needed to determine the empirical relationship between the shapes of conformity functions and normative/informational conformity. Furthermore, future work could explore questions such as: What happens if the information that individual learners receive is not independent, for example when several learners receive information about available options from the same news source? How does the structure of communication networks, such as the presence of echo chambers, affect the outcome? What happens when there are more than two options? Besides social learning, what happens to the system when uninformed decision-makers are more susceptible to stereotypes, such as gender, race, and physical attractiveness \cite{riggle1992bases, mcdermott1998race}? How does the effect of social learners change in multi-stage decision making processes, such as primary and main elections? We hope our work offers an innovative mathematical framework to study the effect of social learning in these more complex scenarios.  \\

\section*{Acknowledgments}
We thank Sidney Redner for helpful discussions and Albert Kao for constructive feedback on an earlier version of the manuscript. VCY was supported by the Santa Fe Institute Omidyar Fellowship, Suzanne Hurst and Samuel Peters. MG was partially supported by NSF Grant 1757211. The funders had no role in study design, data collection and analysis, decision to publish or preparation of the manuscript. Any opinions, findings and conclusions or recommendations expressed in this material are those of the authors and do not necessarily reflect the views of the funders.

\end{document}